\documentstyle[preprint,aps]{revtex}
\begin{document}
\draft
\tighten
\title{Effect   of  friction  on  disoriented  chiral  condensate
formation.}
\author{\bf A. K. Chaudhuri \cite{byline}}
\address{ Variable Energy Cyclotron Centre\\
1/AF,Bidhan Nagar, Calcutta - 700 064\\}

\maketitle

\begin{abstract}

We  have  investigated  the  effect of friction on the DCC domain
formation. We solve the Newton equation of motion  for  the  O(4)
fields,  with  quenched initial condition. The initial fields are
randomly distributed in  a  Gaussian  form.  In  one  dimensional
expansion,  on  the average, large DCC domains can not be formed.
However, in  some  particular  orbits,  large  instabilities  may
occur.   This   possibility  also  greatly  diminishes  with  the
introduction of friction. But, if  the  friction  is  large,  the
system  may  be  overdamped  and  then, there is a possibility of
large DCC domain formation in some events.

\end{abstract}


The possibility of forming disoriented chiral condensate (DCC) in
relativistic  heavy  ion  collisions  has  generated considerable
research activities in recent years. The idea was first  proposed
by Rajagopal and Wilczek \cite{ra93,ra93a,ra95,wi93}. They argued
that  for  a  second  order  chiral  phase transition, the chiral
condensate   can   become   temporarily   disoriented   in    the
nonequilibrium conditions encountered in heavy ion collisions. As
the  temperature drops below $T_c$, the chiral symmetry begins to
break  by  developing  domains  in  which  the  chiral  field  is
misaligned  from its true vacuum value. The misaligned condensate
has the same quark content and quantum numbers as  do  pions  and
essentially  constitute  a  classical pion field. The system will
finally relaxes to the true vacuum and in the  process  can  emit
coherent  pions.  Since the disoriented domains have well defined
isospin orientation,  the  associated  pions  can  exhibit  novel
centauro-like  \cite{la80,ar83,al86,re88} fluctuations of neutral
and charged pions \cite{an89,an89a,bj93,bl92}.

Most dynamical studies of DCC have been based on the linear sigma
model,  in  which  the chiral degrees of freedom are described by
the real O(4)  field  $\Phi=(\sigma,\roarrow{\Pi})$,  having  the
equation of motion,

\begin{equation}
[\Box + \lambda (\phi^2-v^2)]\Phi=H n_\sigma
\label{1}
\end{equation}

The  parameters  of the model can be fixed by specifying the pion
decay constant, $f_\pi$=92 MeV and the meson masses,  $m_\pi$=135
MeV       and       $m_\sigma$=600      MeV,      leading      to
$\lambda=(m_\sigma^2-m_\pi^2)/2f_\pi^2$=20.14                 and
$v=[(m_\sigma^2-3            m_\pi^2)/(m_\sigma^2-m_\pi^2)]^{1/2}
f_\pi$=86.71  MeV  and  $H=(120.55  MeV)^3$  \cite{ra96}.  It  is
apparent  from  eq.\ref{1}  that  the  vacuum  is  aligned in the
$\sigma$  direction  $\Phi_{vac}=(f_\pi,\bf{0})$   and   at   low
temperature  the  fluctuations represent nearly free $\sigma$ and
$\pi$ mesons. At very high temperature well above $v$, the  field
fluctuations are centered near zero and approximate O(4) symmetry
prevails.

It is instructive to decompose the chiral field,

\begin{equation}
\Phi(r,t)=<\phi(t)>+\delta\phi(r,t) \label{2}
\end{equation}

\noindent where $<\phi> $is the average over a suitable region of
space  and can be identified with the (local) order parameter and
$\delta \phi$ are  the  semiclassical  fluctuations  and  can  be
identified  with quasi-particle excitations. Using eq.\ref{2} and
taking the average of eq.\ref{1}, the equation of motion for  the
mean fields in the Hartree approximation can be obtained as,

\begin{equation}
\frac{\partial^2 <\phi>}{\partial t^2} = \lambda(v^2-<\phi>^2-
3<\delta  \phi^2_\|>-<\delta  \phi^2_\bot>)  <\phi>  +H  n_\sigma
\label{3}
\end{equation}

\noindent   where  $\delta  \phi_\|$  is  the  component  of  the
fluctuation parallel to $<\phi>$ and $\delta  \phi_\bot$  is  the
orthogonal  component. This equation imply that the motion of the
mean field is determined by the effective potential,

\begin{equation}
V(<\phi>)=\frac{\lambda}{4}
(<\phi>^2+ 3<\delta \phi^2_\|>+<\delta \phi^2_\bot> -v^2)
\label{4}
\end{equation}

\noindent  which clearly differs from the zero temperature one in
presence of fluctuations. By  varying  the  fluctuations,  chiral
symmetry  can  be  restored  or  spontaneously broken. It is also
evident that the evolution  of  the  order  parameter  critically
depends on the initial values of the fluctuations. When $\delta^2
\equiv  (3<\delta  \phi^2_\|>-<\delta  \phi^2_\bot>)/6$  is large
enough the  chiral  symmetry  is  approximately  (as  H$\neq$  0)
restored.  If  the  explicit  chiral  symmetry  breaking  term is
neglected, the phase  transition  takes  place  at  the  critical
fluctuations    $\delta^2_c\equiv   v^2/6$.   For   $\delta^2   <
\delta^2_c$, the effective potential takes its minimum  value  at
$<\phi>=(\sigma_e,0)$,  where  $\sigma_e$  depends on $\delta^2$.
When the mean fields are displaced from this equilibrium point to
the central  lump  of  the  Mexican  hat  ($<\phi>\sim  0$),  the
effective mass square

\begin{equation}
m^2=\lambda(v^2-<\phi>^2-3<\delta \phi^2_\|>-<\delta
\phi^2_\bot>)
\end{equation}

\noindent will become negative and DCC can form. Since the domain
size  is  directly  related  to  the time scale, during which the
effective mass remains  negative,  it  strongly  depends  on  the
initial  condition, $\delta^2$ and $<\phi>$ of the system. In the
one loop effective theory, the fluctuations can  be  replaced  by
($T^2/2$), their counterpart in a finite temperature field theory
\cite{ga94,ra97}.

Recently,  Biro  and  Greiner  \cite{bi97},  using  the  Langevin
equation for the linear sigma model, investigated  the  interplay
of  friction  and  white  noise  on  the  evolution  of the order
parameter. In the $\phi^4$ theory, it  can  be  shown  explicitly
that  if the hard modes are integrated out on the two loop level,
Langevin type of equation emerges for  the  propagation  of  long
wavelength  fields  \cite{gr97}.  They  found  that  in different
realistic initial volumes ranging from 1 to  1000  $fm^{3}$,  the
average  evolution do not shows any sensible instability. However
individual  events   sometime   shows   significant   growth   of
fluctuations.

In the present paper, we will investigate the role of friction on
the  possible  DCC  domain  formation.  To  this  end,  we  solve
eq.\ref{3}, augmented by a Bjorken type of scaling expansion  and
Rayleigh dissipation term \cite{ra96a}.

\begin{equation}
\frac{\partial^2 <\phi>}{\partial \tau^2}
+(\frac{D}{\tau}+\eta)
\frac{\partial <\phi>}{\partial \tau}
= \lambda[v^2-<\phi>^2-
T^2/2] <\phi> +H n_\sigma
\label{5}
\end{equation}

In the weak coupling limit, the friction $\eta$ is related to the
on-shell plasmon damping rate, $\eta \equiv 2\gamma_{pl}$. In the
standard  $\phi^4$  theory,  the  plasmon  damping  rate  can  be
calculated   for   the   $\sigma$   and    the    $\Pi$    fields
\cite{gr97,pa92}.  Assuming  that  the meson masses are the same,
the friction coefficient can be obtained as \cite{bi97},

\begin{equation}
\eta=2\gamma_{pl}=\frac{9}{16\pi^3}\lambda^2        \frac{T^2}{m}
f_{Sp}(1-e^{-m/T})
\end{equation}

\noindent  where  $f_{Sp}=-\int^x_1  dt \frac{\ln t}{t-1}$ is the
Spence function. At  $T=T_c=\sqrt{2f^2_\pi-2m^2_\pi/\lambda}$=123
MeV,  and  if  $m/T  \simeq 1$, the friction $\eta= 2.2 fm^{-1}$.
This friction may be too  high.  Therefore  we  also  investigate
scenarios  with  1/2  and  1/4  of its value. We also neglect the
temperature dependence of the friction coefficient.

We will solve the eq.\ref{5} with quenched initial condition, for
one  dimensional (D=1) expansion. As argued before, DCC formation
depend critically  on  the  initial  condition.  To  reflect  the
uncertainty  in  the  initial  condition,  the initial fields are
randomly distributed  to  a  Gaussian  form  with  the  following
parameters,

\begin{eqnarray}
<\phi>=&&0\\
<\phi^2>-<\phi>^2=&& v^2/6\\
< \overcirc{\phi}>=&& 0\\
<\overcirc{\phi}^2>-<\overcirc{\phi}>^2=&&4 v^2/6
\end{eqnarray}

For  each  trajectories,  we  compute the characteristic quantity
$m_t=sgn(m^2)\sqrt{|m|}$. The phenomenon of long  wavelength  DCC
amplification will occur whenever this characteristic quantity is
negative. We also calculate the amplification coefficient defined
as,

\begin{equation}
G=\int  |m| \Theta(-m^2) dt
\end{equation}

Amplification of unstable orbits  are  $\sim  exp(G)$.

In fig.1, we have shown the ensemble average of 1000 trajectories
of  the  characteristic  quantity  $m_t$  (solid  line).  We have
considered four different values for  the  friction  coefficient,
$\eta$=0,0.5,1.1  and  2.2 $fm^{-1}$. Also shown is the $m_t$ for
the  most  unstable  orbit  i.e.  the   orbit   for   which   the
amplification  coefficient  G  is  maximum (the dashed line). For
$\eta$=0, $<m_t>$ execute oscillation about its relaxed value  of
140  MeV.  It  remain  positive  throughout  the  evolution. This
indicate that in one dimensional expansion, on  the  average  DCC
like phenomena will not occur. The most unstable orbit shows nice
oscillations   with   slowly  diminishing  amplitude,  indicating
possibility of large DCC  domain  formation  in  this  particular
event.  Thus  while on the average, one dimensional system do not
exhibit DCC  phenomena,  in  some  particular  event,  large  DCC
formation  can  occur.  These  results  are in agreement with the
other calculations \cite{ra96b,bi97}. With  the  introduction  of
friction  the  results  changes.  For  small  friction $\eta$=0.5
$fm^{-1}$, $<m_t>$ do not oscillate much. It also remain positive
throughout the evolution. Like the  previous  frictionless  case,
DCC  phenomena  is  not  expected  on the average level. The most
unstable orbit exhibit rapidly damping  oscillation.  It  remains
negative but for a short period of time. As size of DCC domain is
directly  related to the time during which $m_t$ remain negative,
the result indicate that even for the most unstable orbit,  large
DCC  domains  can not be expected in this situation. Introduction
of friction greatly  diminishes  the  possibility  of  large  DCC
domain  formation.  With  further increase of the friction result
changes again. With $\eta$=1.1 $fm^{-1}$,  $<m_t>$  do  not  show
oscillation.  But we find that it enters into the unstable region
for a short duration of time. With moderately large friction, DCC
phenomena can occur on average  level  also.  The  most  unstable
orbit shows a interesting behavior. Once $m_t$ reaches a negative
value,  it remains negative for a long duration, then relaxing to
its stable value  of  140  MeV.  Similar  behavior  is  seen  for
$\eta=2.2  fm^{-1}$.  $m_t$  for  the  most  unstable orbit stays
negative  for  quite  a  long  time  ($\sim$15  fm),   indicating
possibility  of  a  large DCC formation in this particular event.
The results indicate that if the friction is sufficiently  large,
one  can hope to form large DCC domains in some particular event.
This result can be understood  easily.  With  large  $\eta$,  the
system  is  overdamped  and  once the system enters into unstable
regime, it can not come out from it and remains unstable for long
duration.

In  order  to  check that the oscillations in $m_t$ are stable or
not, we have performed the  time  series  analysis  according  to
Hurst  \cite{hu51,hu91} and obtained Hurst coefficient (H). Hurst
coefficient varies between 0-1. For a perfectly random  data  set
it  is  0.5.  If $H>0.5$, the data set is persistent and they are
antipersistent if $H<.5$. For the persistent  data  set,  if  the
trend  or  behaviour  in the data set is increasing or decreasing
over a certain  unit  interval  of  time,  it  will  continue  to
increase  or  decrease over such an interval, but no such comment
can be made about antipersistent orbits. The most unstable orbits
shown in fig.1 are found to be persistent ($H>.5$). In fig.2,  we
have shown the distribution of the Hurst coefficient for the 1000
trajectories  calculated.  For  the  frictionless system, quite a
large number of trajectories are found to be antipersistent.  The
oscillation  for  these  orbits  may  eventually  die  out.  With
increasing friction, this number decreases  and  with  $\eta$=2.2
$fm^{-1}$, we find all the trajectories are persistent.

The  amplification  coefficient  (G) is an important parameter in
the theoretical analysis of DCC. Larger the G, larger is the  DCC
domain  size.  In  fig.3, we have shown the distribution of G for
all the 1000 trajectories as obtained in the present calculation.
The distribution of the  persistent  trajectories  ($H>.5$),  are
also  shown  (the  shaded  region).  It  is  evident that maximum
unstability is obtained in the frictionless case, with G as large
as 2.4. Frictionless case also shows marked difference when  only
the  persistent  trajectories  are considered. For the persistent
trajectories the distribution decreases continuously with  G,  in
contrast, when all the trajectories are considered, the number of
trajectories  remains  nearly  constant  between  G=.5-1.5.  With
friction, as most of the trajectories are  persistent,  there  is
not  much  difference  between  the  total  distribution  and the
distribution of persistent trajectories.  As  is  the  case  with
frictionless  case, we find that number of trajectories decreases
continuously with increasing G.

The  pions  from  a  DCC domain shows anomalous behavior, as they
have definite isospin orientation. Defining the neutral to  total
pion ratio $f$

\begin{equation}
f=\frac{\pi^0}{\pi^+ +\pi^- +\pi^0}
\end{equation}

the  probability  to  obtain a particular fraction $f$ in case of
DCC formation can be obtained as \cite{an89,bj93,bl92},

\begin{equation}
P(f)=\frac{1}{2\sqrt{f}} \label{6}
\end{equation}

incontrast  to  the  binomial  distribution  peaked  at $1/3$ for
normal  hadronic  reaction.  From  eq.\ref{6},   an   equivalent
probability that $f<f_1$ can be obtained,

\begin{equation}
P(f<f_1)=\sqrt{f} \label{7}
\end{equation}

Noting  that  the  pion yield will be proportional to $\int \Pi^2
\tau d\tau$, we have calculated this probability. The results are
shown in fig.4 (solid line). The dashed lines are the probability
distribution given by eq.\ref{7}. While for frictionless and for
small friction, the agreement between the two are reasonable, for
moderate to high friction, they in very good agreement. It  seems
that  the  pions  are  all  from  DCC  like events. This raises a
question. Preceding analysis shows that it  is  not  possible  to
obtain  large  DCC  domain  except  in a few trajectories. Yet we
obtain neutral to total  pion  ratio  following  the  probability
distribution of a DCC domain. This indicate that $1/\sqrt{f}$ may
not be good signal to identify the pions from DCC domains.

To  conclude,  we have investigated the effect of friction on the
possible DCC domain formation. It was shown  that  while  in  one
dimensional case, on average DCC formation is not possible, there
is  finite  probability  for  DCC  domain formation in a event by
event basis. We also find that friction  strongly  inhibit  large
DCC  formation  even  in  a event by event basis. However, if the
friction coefficient is sufficiently large, the system may become
overdamped and then large DCC domain can be formed.

\begin{figure}
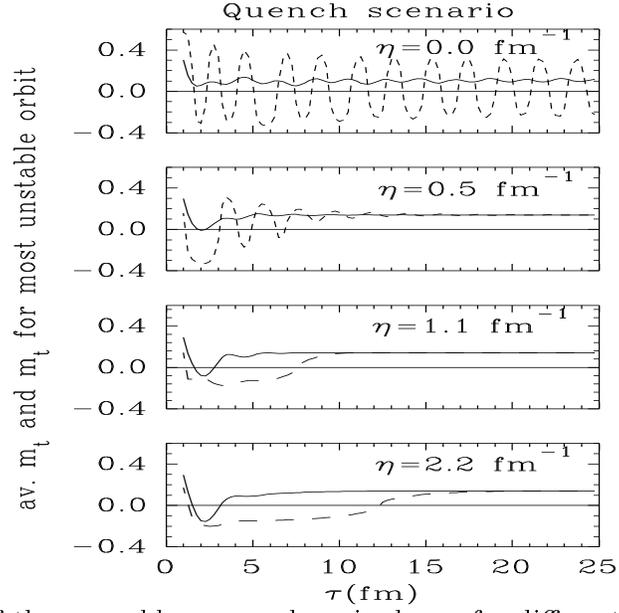

\caption{Evolution  of  the  ensemble averaged $m_t$ is shown for
different  friction  coefficients  (the  solid  line).  The  most
unstable  orbit,  for  which the amplification coefficient is the
largest is also shown (the dashed line).}
\end{figure}
\begin{figure}
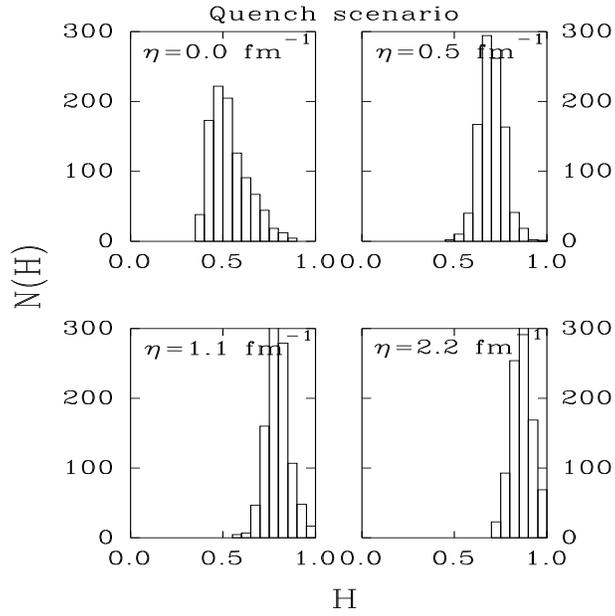

\caption{Histogram  showing the distribution of Hurst coefficient
for the 1000 trajectories.}
\end{figure}
\begin{figure}
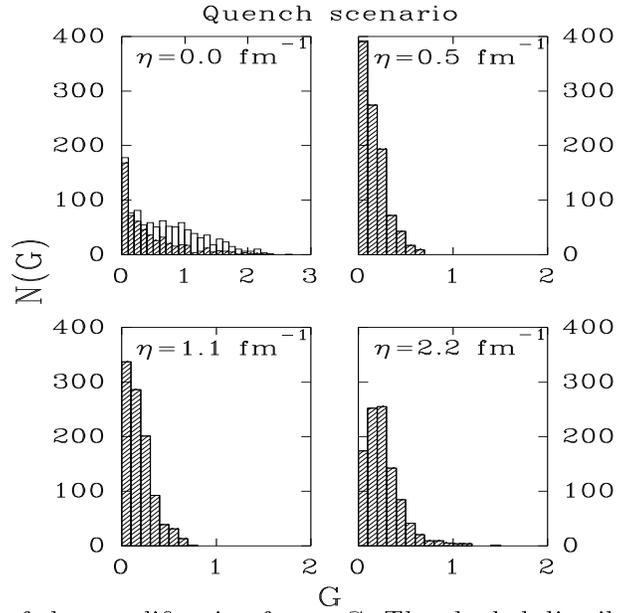

\caption{Distribution of the amplification factor G.  The  shaded
distribution  is  for  the  persisten  trajectories  with   Hurst
coefficient $H>0.5$.}
\end{figure}
\begin{figure}
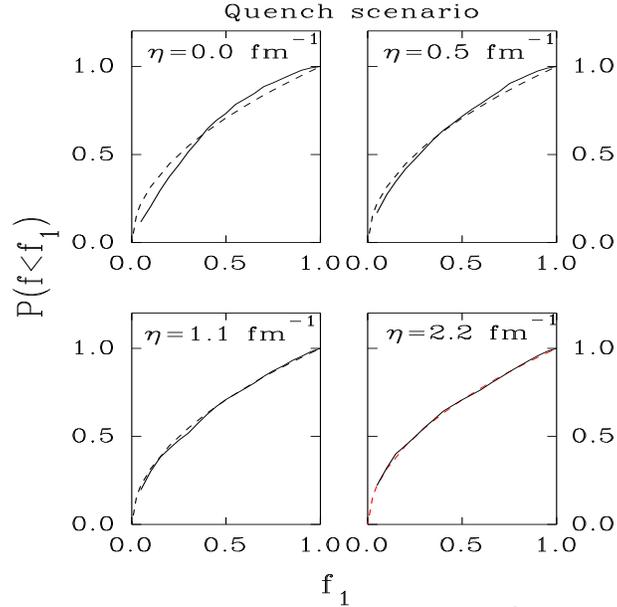

\caption{ The probability that the neutral to total pion ratio $f$
is  less  than  $f_1$. The solid line is the calculated while the
dashed line is from eq.\protect{\ref{7}}.}
\end{figure}
\end{document}